\documentclass[twocolumn,prl,a4paper,aps,showpacs,amsmath,amssymb]{revtex4}
\usepackage{graphicx}
\usepackage{dcolumn}
\usepackage{bm}

\begin{document}

\title{Tunable dipolar magnetism in high-spin molecular clusters}

\author{M. Evangelisti$^{1,*}$, A. Candini$^{1,2}$, A. Ghirri$^{1,2}$, M. Affronte$^{1,2}$, G. W.
Powell$^{3}$, I. A. Gass$^{4}$, P. A. Wood$^{4}$, S. Parsons$^{4}$, E. K. Brechin$^{4}$, D. Collison$^{3}$,
and S. L. Heath$^{3}$}

\affiliation{$^{1}$ National Research Center on ``nanoStructures and bioSystems at Surfaces'' (S$^{3}$),
INFM-CNR, 41100 Modena, Italy\\ $^{2}$ Dipartimento di Fisica, Universit\`{a} di Modena e Reggio Emilia,
41100 Modena, Italy\\ $^{3}$ Department of Chemistry, University of Manchester, M13 9PL Manchester, United Kingdom\\
$^{4}$ School of Chemistry, University of Edinburgh, EH9 3JJ Edinburgh, United Kingdom}
\date{\today}

\begin{abstract}
We report on the Fe$_{17}$ high-spin molecular cluster and show that this system is an exemplification of
nanostructured dipolar magnetism. Each Fe$_{17}$ molecule, with spin $S=35/2$ and axial anisotropy as small
as $D\simeq -0.02$~K, is the magnetic unit that can be chemically arranged in different packing crystals
whilst preserving both spin ground-state and anisotropy. For every configuration, molecular spins are
correlated only by dipolar interactions. The ensuing interplay between dipolar energy and anisotropy gives
rise to macroscopic behaviors ranging from superparamagnetism to long-range magnetic order at temperatures
below 1~K.
\end{abstract}

\maketitle

A rejuvenated interest in phase transitions driven only by dipolar interactions has emerged since the
experimental discovery that the magnetic molecular materials Cr$_{4}$~\cite{Bino88} and
Mn$_{6}$~\cite{Morello03,Morello06} provide attractive examples of pure dipolar magnets. These are
nanostructured such that molecular clusters replace what atoms are to conventional materials.
Quantum-mechanical superexchange interactions within each molecule result in net (high-)spin values per
molecule at low temperatures. In parallel, dipolar interactions provide the only source of coupling between
the molecular spins arranged in crystallographic lattices. Assuming each molecule as a high-spin point-like
dipole, the macroscopic properties of dipolar magnets can be precisely predicted because dipole-dipole
interactions are calculated without involving any adjustable
parameter~\cite{Morello06,Panissod02,Fernandez00,Fernandez02,Chudnovsky01,Fernandezun}. These ideal materials
are however very difficult to obtain. As often is the case, intermolecular superexchange interactions may not
be negligible at very low temperatures where long-range magnetic order (LRMO) takes place. The consequence is
that correlations between the molecules are often established by quantum-mechanical superexchange
interactions at short ranges, whose macroscopic prediction is made difficult by their strong dependence on
electronic details. Indeed, intermolecular superexchange interactions were found to be responsible for the
observed LRMO in the high-spin molecular clusters Fe$_{19}$~\cite{Affronte02}, Mn$_{4}$Br~\cite{Yamaguchi02},
Mn$_{4}$Me~\cite{Evangelisti04}, and Fe$_{14}$~\cite{Evangelisti05}, while they likely play a mayor role also
in Mn$_{12}$~\cite{Luis05}.

The absence of any superexchange pathway between the molecules is not the only prerequisite needed for the
experimental observation of dipolar order. An obvious requirement is that molecules should have large
molecular spins to lead to accessible ordering temperatures. Another complication is added by the cluster
magnetic anisotropy. Crystal-field effects give rise to anisotropy energy barriers for each molecule that
result in slow magnetic relaxation below a certain blocking temperature. The cluster anisotropy energies
favor the molecular spin alignment along dictated directions, thus competing with the intermolecular
coupling. The anisotropy therefore has to be very small, such that the spin-lattice relaxation is kept
sufficiently fast down to temperatures low enough for LRMO to be observed~\cite{Morello03,Evangelisti04}.

\begin{figure}[b!]
\centering{\includegraphics[angle=0,width=6.8cm]{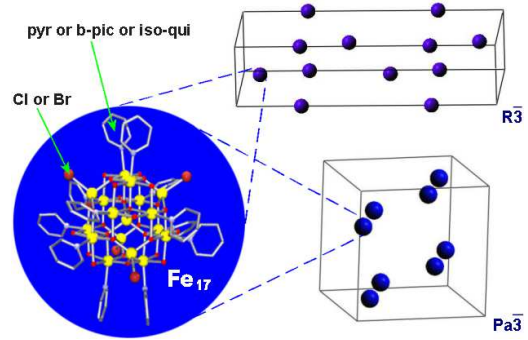}} \caption{(color online) The Fe$_{17}$ molecule
containing 17 magnetically coupled Fe$^{3+}$ atoms (Fe = yellow balls; Cl or Br = red balls ; O = small red
balls), together with the packing in two different unit cells: $R\bar{3}$ (trigonal) and $Pa\bar{3}$
(cubic)~\cite{note2}.}
\end{figure}

In this Letter, we present the Fe$_{17}$ molecular nanomagnet~\cite{Powell04}, containing 17 Fe$^{3+}$ atoms
per molecule linked via oxygen atoms (Fig.~1). Carrying very-large spin $S=35/2$ and axial anisotropy as
small as $D\simeq -0.02$~K, the Fe$_{17}$ high-spin molecular cluster represents an excellent candidate for
these studies. In addition, these molecules are only bound together in the crystal by van der Waals forces
thus prohibiting any intermolecular superexchange pathway. What makes Fe$_{17}$ a {\it unique} model system
is that we are able, by changing the crystallization conditions, to change the molecular packing {\it
without} affecting the individual molecules, that is keeping the surrounding ligands, the molecular high-spin
ground-state and magnetic anisotropy unaltered. In other words, we succeed for the first time in efficiently
tuning the dipolar coupling between molecules with respect to the single-molecule properties. The resulting
interplay gives rise to macroscopic behaviors ranging from superparamagnetic blocking to LRMO.

\begin{figure}[b!]
\centering{\includegraphics[angle=0,width=7.6cm]{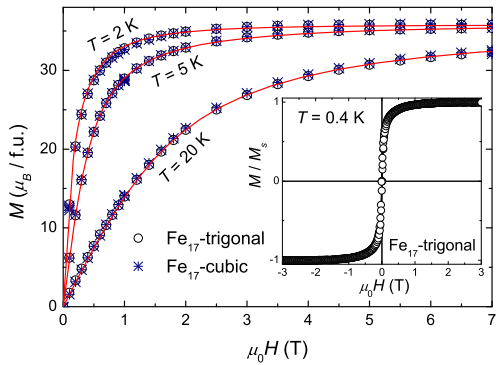}} \caption{(color online) Isothermal molecular
magnetization for both Fe$_{17}$-trigonal ($\circ$) and Fe$_{17}$-cubic ($\ast$) collected at $T=2, 5$ and
20~K. Solid lines are the results of the fit (see text), yielding net molecular spin $S=35/2$ and axial
$D=-0.023$~K. Inset: Hysteresis loop of Fe$_{17}$-trigonal at $T=0.4$~K.}
\end{figure}

The Fe$_{17}$ molecules are obtained by dissolving either FeBr$_{3}$ or FeCl$_{3}$ in a coordinating base,
e.g. pyridine (pyr), beta-picoline (b-pic) or iso-quinoline (iso-qui) that also acts as solvent. To
crystallize the product (Fe$_{17}$), we slowly diffuse a second (often non-coordinating) co-solvent like
diethyl-ether (Et$_{2}$O), acetone (Me$_{2}$CO), acetonitrile (MeCN), iso-propylalcohol (IPA), {\it etc.},
into the basic solution. The product is generally soluble in the first solvent (e.g. pyr) but insoluble in
the second (e.g. Et$_{2}$O) and by slowly diffusing the second solvent in, we crystallize the product. In
this way we obtain several derivatives having the same Fe$_{17}$ magnetic core~\cite{note1}. Whilst the spin
value is preserved throughout the whole Fe$_{17}$ family, the anisotropy may change significantly. We have
synthesized a number of new Fe$_{17}$ clusters containing bromide ions in which we can either (i) exchange
the pyr ligands for b-pic or iso-qui ligands (Fig.~1) thus modifying only the outer organic coating of the
Fe$_{17}$, such that the major change is purely intramolecular ({\it anisotropy}); or (ii) change the
crystallizing co-solvent such that we change only the packing ({\it space group}) of the molecules in the
crystal. For example, the reaction between FeBr$_{3}$ and pyr in the presence of Me$_{2}$CO affords the
Fe$_{17}$ magnetic core crystallized in the trigonal space group $R\bar{3}$, whilst the same reaction but in
the presence of IPA gives an identical Fe$_{17}$ magnetic core but crystallized in the cubic space group
$Pa\bar{3}$ (Fig.~1). By defining the organic ligand and subsequent crystallizing conditions, we can
therefore reproducibly generate different arrays of this molecular magnet. In what follows, we focus on the
above-mentioned Br derivatives of the Fe$_{17}$ molecule having trigonal or cubic symmetries~\cite{note2}.
Measurements of magnetization down to 2~K and specific heat down to $\approx 0.3$~K on powder samples, were
carried out for the $0<H<7$~T magnetic field range. Magnetization, susceptibility and relaxation measurements
below 2~K were performed using home-made Hall microprobes. In this case, the grain-like samples consisted of
collections of small crystallites of c.a. $10^{-3}$~mm$^{3}$. For measurements performed on powder samples,
the calculated fits were obtained taking into account spin random orientations.

Field-dependencies of the molar magnetization $M(H)$ for both Fe$_{17}$-trigonal and Fe$_{17}$-cubic were
collected for $T=2, 5$ and 20~K (Fig.~2). The important result is that the $M(H)$ curves depend on the
applied-field in an {\it identical} manner regardless of the trigonal or cubic symmetry. This implies that
the Fe$_{17}$ magnetic molecule (that is the spin ground-state and anisotropy) is the same in both complexes.
If we consider the single-spin Hamiltonian
$\mathcal{H}=DS_{z}^{2}+g\mu_{B}\overrightarrow{H}\cdot\overrightarrow{S}$, the magnetization in the whole
field-range can be well fitted with net molecular spin $S=35/2$, zero-field splitting $D=-0.023$~K and
$g=2.06$ for both complexes. Although smaller trigonal components could be present, the data do not justify a
more sophisticated fitting.

\begin{figure}[b!]
\centering{\includegraphics[angle=0,width=7.6cm]{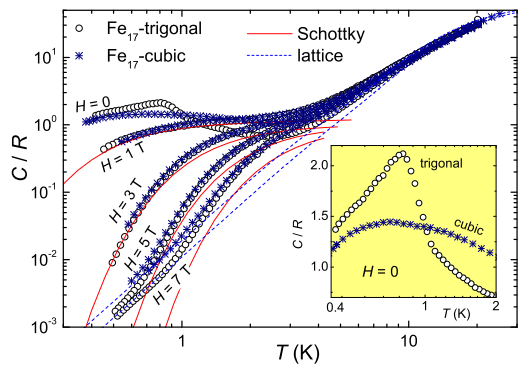}} \caption{(color online) Specific heat of
Fe$_{17}$-trigonal and Fe$_{17}$-cubic for several applied-fields, as labeled. Drawn curves are explained in
the text. Inset: Magnification of the low-$T$ / zero-field range showing the different ordering behaviors.}
\end{figure}

Figure~3 shows the collected specific heat $C(T,H)$ data of both Fe$_{17}$ molecular compounds as function of
temperature for several applied-fields. At first sight and as for the $M(H)$ data (Fig.~2), the $C(T,H)$ of
Fe$_{17}$-trigonal does not differ from that of Fe$_{17}$-cubic, at least for $H>0$. The main difference is
in the zero-applied field data for which a $\lambda$-type anomaly centered at $T_{C}=0.81$~K is observed for
trigonal symmetry (inset of Fig.~3). Anticipating the discussion below, this feature reveals the onset of
LRMO, the magnetic nature is indeed proven by its disappearance upon application of $H$. Clearly, the
$\lambda$-type anomaly arises on top of a much broader one, which shifts with increasing applied-field
towards higher temperatures. Because of the small anisotropy ($D\simeq -0.02$~K), it is expected that the
magnetic contribution to $C(T,H)$ for $H\geq 1$~T is due to Schottky-like Zeeman splitting of the otherwise
nearly degenerate energy spin states. Indeed, the calculated Schottky curves (solid lines in Fig.~3) arising
from the field-split levels accounts very well for the experimental data. The same behavior is followed by
Fe$_{17}$-cubic except that no sign of LRMO is apparently observed.

As particularly evident in the low-$T$ / high-$H$ region of Fig.~3, phonon modes of the crystal lattice
contribute differently to $C(T)$ of Fe$_{17}$-trigonal and Fe$_{17}$-cubic. We estimated the lattice
contributions (dashed lines in Fig.~3) by fitting to a model given by the sum of a Debye term for the
acoustic low-energy phonon modes plus an Einstein term that likely arises from intramolecular vibration
modes. From the field-dependencies of $M(T,H)$ and $C(T,H)$, we have already deduced that the individual
Fe$_{17}$ molecule remains identical regardless of space group. Therefore, it is not surprising that the fit
provides the same Einstein temperature $\theta_{E}\simeq 42$~K for both compounds (Fig.~3). Contrary,
low-energy phonon modes result in different Debye temperatures whose values are $\theta_{D}\simeq 23$~K and
28~K for Fe$_{17}$-cubic and Fe$_{17}$-trigonal, respectively. Because Fe$_{17}$-cubic has larger
intermolecular distances~\cite{note2}, softer low-energy modes, yielding smaller $\theta_{D}$, are indeed to
be expected. The so-obtained lattice contributions allow us to estimate the entropy changes $\Delta S$ by
using the relation $\Delta S/R=\int_{0}^{\infty}C_{m}(T)/(RT){\rm d}T$ where $C_{m}(T)$ is the magnetic
contribution obtained from $C(T)$ after subtraction of the respective lattice contribution. For both
compounds, the obtained $\Delta S$ amounts to $3.7~R$, which is in good agreement with the entropy expected
$R~{\rm ln}(2$S$+1)\simeq 3.6~R$, given $S=35/2$. As already anticipated, we can therefore safely attribute
$T_{C}=0.81$~K to the LRMO temperature of the molecular spins in Fe$_{17}$-trigonal.

\begin{figure}[t!]
\centering{\includegraphics[angle=0,width=7.6cm]{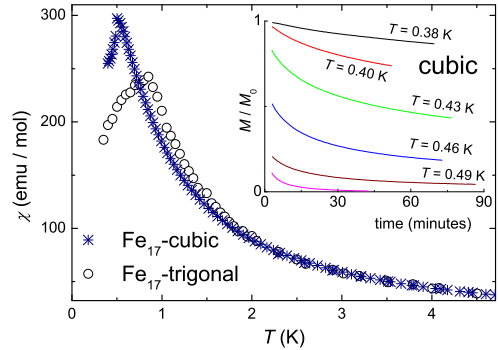}} \caption{(color online) Magnetic susceptibility
of Fe$_{17}$-trigonal and Fe$_{17}$-cubic for $H=0.01$~T. Inset: Time decay of $M$ for Fe$_{17}$-cubic,
measured at zero-applied-field after saturation for the indicated temperatures.}
\end{figure}

Susceptibility measurements (Fig.~4) reveal sharp anomalies that take place at $\sim T_{C}$ for
Fe$_{17}$-trigonal, corroborating the LRMO deduced from specific heat data and at $T_{B}\simeq 0.5$~K for
Fe$_{17}$-cubic, whose nature is discussed below. For $T>4$~K, both susceptibilities tend to overlap each
other (Fig.~4). The observed behavior in Fe$_{17}$-trigonal is compatible with a ferromagnetically ordered
phase, in which demagnetization effects become important. The measured susceptibility at $T_{C}$ is close to
the estimated limit for a ferromagnetic grain-like sample, $\chi_{N}=1/\rho N\simeq 227$~emu/mol (see
Fig.~4), where $\rho=3.32$~g/cm$^{3}$ is the density of Fe$_{17}$-trigonal, and $N=4\pi/3$ is the
demagnetizing factor of the grain-like sample, approximated to a sphere. For the $5~{\rm K}\lesssim T\lesssim
80$~K temperature range, the fit to the Curie-Weiss law $\chi=C/(T-\theta)$ for the susceptibility of
Fe$_{17}$-trigonal corrected for the demagnetizing field, $\chi=\chi\prime/(1-\rho N\chi\prime)$, provides
$C=175.4$~emuK/mol and $\theta=0.9$~K, in agreement with the observed ferromagnetic order at $T_{C}\simeq
0.8$~K. The Curie constant $C$ equals (within error) the expected value of a (super)paramagnet with spin
$S=35/2$ and $g=2.06$, as deduced above from the magnetization data. This analysis is corroborated by the
hysteresis loop we collected in the ordered phase (inset of Fig.~2), for which Fe$_{17}$-trigonal behaves as
a soft ferromagnet with a coercivity of $\sim 60$~Oe. We recall that from $M(H)$ curves, we estimated the
anisotropy $D=-0.023$~K, which likely causes a pinning of the domain-wall motions responsible therefore for
the slow decrease of the experimental susceptibility below $T_{C}$ (Fig.~4).

The occurrence of a sharp peak at $T_{B}\simeq 0.5$~K in the susceptibility of Fe$_{17}$-cubic (Fig.~4) has
apparently no counterpart in the specific heat (inset of Fig.~3). We therefore exclude LRMO as a possible
source. We recall that intermolecular distances for Fe$_{17}$-cubic are slightly larger than that of
Fe$_{17}$-trigonal~\cite{note2}. It is then reasonable to assume that in Fe$_{17}$-cubic the intermolecular
coupling is weaker, and that the molecular anisotropy is the predominant energy. This would lead to
superparamagnetic blocking at $T_{B}$ of the molecular spins along preferred directions dictated by the
anisotropy. To better elucidate this point we performed magnetic relaxation experiments in Fe$_{17}$-cubic at
temperatures below $T_{B}$. We firstly applied a field necessary to saturate the magnetization of the sample
at 2~K. We then cooled down to a given temperature below $T_{B}$ and, upon removing the field, we followed
the relaxation of the Fe$_{17}$ molecules to thermal equilibrium by collecting the time decay of the
magnetization. Results are shown in the inset of Fig.~4, where it is seen that the decay neatly slows down
below $T_{B}$, as expected for a superparamagnet. Magnetization data are well described by a stretched
exponential decay $M/M_{0}={\rm exp}(-t/\tau)^{\beta}$ where $M_{0}$ is the initial magnetization, $\beta$
the stretched parameter and $\tau$ the characteristic decay time. The $T$-dependence of $\tau$ (not presented
here) follows an Arrhenius law providing the activation energy $U=9.0$~K, that given $S=35/2$ and
$U=-D(S^{2}-1/4)$, corresponds to $D\simeq -0.03$~K, which is of the same order of that estimated above. We
note that $U$ of the Fe$_{17}$ molecule is about eight times smaller than that of the well-known
single-molecule magnet Mn$_{12}$-ac~\cite{Sessoli93}. As a result of similar spin dynamics, the same ratio
holds for the respective blocking temperatures as well.

The Fe$_{17}$ molecules are magnetically isolated from each other as evidenced by the large intermolecular
distances, for instance {\it all} Fe-Fe distances are greater than 8.7~\AA~for adjacent molecules in
Fe$_{17}$-trigonal. A close inspection of the crystallography of both materials does not reveal any
intermolecular superexchange pathway nor any evidence of $\pi$-stacking of the pyridine rings~\cite{note2}.
These facts show that the dipolar interaction is solely responsible for the observed macroscopic behaviors.
By switching from trigonal to cubic symmetry we change not only the arrangement and reciprocal distances of
the Fe$_{17}$ molecules, but accordingly also the dipolar interaction energies $E_{dip}$. We performed
extensive calculations of $E_{dip}$ assuming several magnetic configurations of $S=35/2$ point-like dipoles
arranged in analogous crystallographic lattices to that of Fe$_{17}$-trigonal and Fe$_{17}$-cubic. In
particular, the position of the spins was fixed accordingly to molecular centroids. Interestingly, for
$E_{dip}$ we found up to an order of magnitude change by switching from cubic to trigonal symmetry. Since the
distance between nearest neighbors change by less than 10\% by switching from Fe$_{17}$-cubic to
Fe$_{17}$-trigonal~\cite{note2}, one has to conclude that lattice symmetries play the major role in
determining $E_{dip}$. This neatly illustrates that the nature of the magnetic order should not be deduced by
simply comparing the ordering temperature with the interaction energy between a pair of nearest spins.

For Fe$_{17}$-trigonal the magnitude of the calculated $E_{dip}$ does justify that LRMO is here driven by
dipolar coupling between the molecules. We do not have, however, enough evidence to discriminate which
magnetic structure is realistically the most probable one. Among the magnetic structures considered in our
simulations and on basis of our experimental data suggesting a ferromagnetic nature of the ordered phase,
promising candidates seem to be the alignment of the molecular spins along the [100] direction and that along
the [2\={2}1] direction. These configurations have indeed the lowest calculated values ($-E_{dip}\simeq
0.8$~K and 0.6~K, respectively), which are of the correct order with respect to the experimental
$k_{B}T_{C}$. We here anticipate that preliminary neutron diffraction experiments~\cite{neutron} have
recently corroborated the onset of the magnetic phase transition for Fe$_{17}$-trigonal.

Summing up, we have developed a synthetic strategy to prepare (Fe$_{17}$) nanomagnets with varying crystal
symmetry. We experimentally demonstrate that Fe$_{17}$ represents the first molecular system to undergo
either LRMO or superparamagnetic blocking of the molecular spins depending on the symmetry. We show that this
results from the interplay of the dipolar magnetic coupling between the molecular spins, with respect to the
single-molecule magnetic anisotropy. That supramolecular chemistry leads to fascinating ordered arrangements
of identical high-spin nanomagnets is no novelty; that these arrangements can be achieved without affecting
the magnetic properties of the individual nanomagnets (e.g. keeping unaltered the cluster spin ground-state
and magnetic anisotropy) is a step forward in the manipulation of the magnetic interactions at the nanometer
scale. The Fe$_{17}$ system is therefore a test model material for workers interested in the modelization of
phase transitions purely driven by dipolar interactions.

The authors are indebted to F. Luis for useful comments, C. Vecchini, O. Moze, D.H. Ryan, and L.M.D.
Cranswick for the neutron diffraction experiments, and M. Helliwell for the structure analysis. This work is
partially supported by Italian MIUR under FIRB project no. RBNE01YLKN and by the EC-Network of Excellence
``MAGMANet'' (contract No. 515767).


\begin{thebibliography}{99}
\bibitem{note1} For instance, the reaction between FeCl$_{3}$ and pyr in the presence of MeOH produces the complex
Hpyr[Fe$_{17}$O$_{16}$(OH)$_{12}$(pyr)$_{12}$Cl$_{4}$]Cl$_{4}$ whose structure and preliminary magnetic
properties were recently reported in Ref.~\cite{Powell04}, where we showed that it has $S=35/2$ and small
cluster anisotropy.
\end{thebibliography}
\end{document}